\documentclass[3p,times,preprint,twocolumn]{elsarticle}
\biboptions{sort&compress}
\usepackage{enumerate}
\usepackage{mathtools}
\usepackage{color}
\usepackage{xcolor}

\usepackage[normalem]{ulem}

\makeatletter

\usepackage{amssymb,amsmath,amsthm}
\usepackage[title,toc]{appendix}
\usepackage{graphicx}
\usepackage{color}
\usepackage{hyperref}
\theoremstyle{definition}

\begin{document}

\begin{frontmatter}

\title{Primordial bouncing cosmology in the Deser-Woodard nonlocal gravity}

\author[1,2]{Che-Yu Chen}
\ead{b97202056@gmail.com}

\author[1,2,3]{Pisin Chen}
\ead{pisinchen@phys.ntu.edu.tw}

\author[4]{Sohyun Park}
\ead{park@fzu.cz }

\address[1]{Department of Physics and Center for Theoretical Sciences, National Taiwan University, Taipei, Taiwan 10617}
\address[2]{LeCosPA, National Taiwan University, Taipei, Taiwan 10617}
\address[3] {Kavli Institute for Particle Astrophysics and Cosmology, SLAC National Accelerator Laboratory, Stanford University, Stanford, CA 94305, U.S.A.}
\address[4]{CEICO, Institute of Physics of the Czech Academy of Sciences,
Na Slovance 2, 18221 Prague 8 Czech Republic}

\begin{abstract}
The Deser-Woodard (DW) nonlocal gravity model has been proposed in order to describe the late-time acceleration of the universe without introducing dark energy. In this paper we focus, however, on the early stage of the universe and demonstrate how a primordial bounce in the vacuum spacetime can be realized in the framework of the DW nonlocal model. We reconstruct the nonlocal distortion function, which encodes all the modifications to the Einstein-Hilbert action, in order to generate bouncing solutions to solve the initial singularity problem. We show that the initial conditions can be chosen in such a way that the distortion function and its first order derivative approach zero after the bounce and the standard cosmological solution described by general relativity is recovered afterwards. We also study the evolution of anisotropies near the bounce. It turns out that the shear density defined by the anisotropy grows towards the bounce, but due to the presence of nonlocal effects, it grows in a milder manner compared with that in Einstein gravity.
\end{abstract}

\begin{keyword}
Modified theories of gravity \sep
Bouncing cosmology \sep
Nonlocal gravity

\end{keyword}
\end{frontmatter}

\section{Introduction}
One of the fundamental problems in Einstein's general relativity (GR) is the existence of spacetime singularities associated with the energy conditions for matter fields. The definition of spacetime singularities in the notion of geodesic incompleteness in GR has been established by Hawking and Penrose \cite{Hawking:1969sw}. In the standard big bang cosmology, the universe started with a big bang singularity, at which the spacetime curvature diverges, jeopardizing the applicability of GR at the initial time.   

It has been suggested that the big bang singularity can be replaced by a nonsingular bouncing scenario in which all curvature invariants become finite. In fact, such bouncing cosmologies have been realized in the context of loop quantum cosmology \cite{Bojowald:2001xe,Ashtekar:2011ni} and string theory \cite{Veneziano:1991ek}. In practice, one can either construct bouncing solutions within the framework of GR by introducing matter fields which violate null energy conditions \cite{Cai:2007qw,Cai:2009zp,Easson:2011zy}, or modify GR such that bouncing solutions can be obtained without violating any energy conditions \cite{Mukhanov:1991zn,Cai:2011tc,Koshelev:2012qn,Biswas:2012bp,Cai:2012ag,Bamba:2013fha,Nojiri:2014zqa,Saridakis:2018fth,Banados:2010ix,Bouhmadi-Lopez:2017lbx}. See Ref.~\cite{Novello:2008ra} and the references therein for a review.

A nonlocal addition to GR is motivated by the fact that quantum loops of massless particles such as gravitons inevitably give rise to nonlocal quantum corrections, see for example \cite{Donoghue:2014yha}. It has been noted that a nonlocal quantum effective action might be obtained from a fundamental local Lagrangian through the gravitational vacuum polarization of infrared gravitons \cite{Woodard:2014iga,Woodard:2018gfj}. However, such a derivation from first principles is not yet available, therefore a compromised approach has been to construct a nonlocal effective action with an aim to describe phenomena that cannot be explained by GR with ordinary matter. Nonlocal models replacing dark matter, sometimes called the nonlocal metric realization of MOND (MOdified Newtonian Dynamics), have also been proposed and studied in Ref.~\cite{Soussa:2003vv,Deffayet:2011sk,Woodard:2014wia,Deffayet:2014lba,Kim:2016nnd,Tan:2018bfp,Arraut:2013qra}. For the nonlocal models that attempt to explain inflation, see Refs.~\cite{Tsamis:1997rk,Tsamis:2009ja,Tsamis:2010pt,Tsamis:2014hra,Koshelev:2016vhi,Tsamis:2016boj}.

So far the most popular application of nonlocal gravity has been the late-time cosmic acceleration without dark energy \cite{Wetterich:1997bz,Deser:2007jk, Barvinsky:2003kg,Barvinsky:2011hd,Barvinsky:2011rk,Maggiore:2013mea,Maggiore:2014sia,Vardanyan:2017kal,Amendola:2017qge,Tian:2018bmn}. Among such models, the two most studied are the Deser-Woodard (DW) model \cite{Deser:2007jk}, which distorts the Ricci scalar $R$ by a function of the nonlocal dimensionless scalar $\square^{-1}R$, and the Maggiore-Mancarella (MM) model \cite{Maggiore:2014sia}, which introduces a mass parameter $m^2$ multiplied by the dimensionful nonlocal scalar $\square^{-2}R$. For the DW model, cosmological perturbations and structure formation  have been studied  in Refs.~\cite{Park:2012cp, Dodelson:2013sma, Park:2016jym, Nersisyan:2017mgj, Park:2017zls, Amendola:2019fhc}. The issues concerning causality, localization, degrees of freedom and stability \cite{Zhang:2016ykx, Nojiri:2007uq, Nojiri:2010pw, Zhang:2011uv, Park:2019btx}, and the gravitational energy-momentum flux due to an isolated system \cite{Chu:2018mld} have been also discussed. The phenomenology of the MM model has also been extensively studied \cite{Dirian:2014ara, Barreira:2014kra, Dirian:2014xoa, Dirian:2014bma, Dirian:2016puz, Nersisyan:2016hjh, Dirian:2017pwp,Kumar:2018pkb}.  

However, a very recent analysis using the constraint from Lunar Laser Ranging has ruled out both the DW and MM models \cite{Belgacem:2018wtb}. For the case of the DW model, this is essentially because the assumption that the nonlocal scalar has opposite signs ($\square^{-1}R < 0$ for the cosmological scales and $\square^{-1}R >0$ for the solar system scales) turns out to be wrong. It was shown that $\square^{-1}R$ remains negative inside gravitationally bound systems because of an integration constant required in this case \cite{Belgacem:2018wtb}. Deser and Woodard have proposed an improved model based on another dimensionless nonlocal scalar made out of $\square^{-1}R$ \cite{Deser:2019lmm}. While waiting for more detailed analyses for the new DW model, in this paper we consider a completely different application of the original DW model, namely the primordial bouncing cosmology. 

The rest of this paper is organized as follows. In Section~\ref{sec.2}, we review the action and equations of motion for the DW model. Section~\ref{sec.3} newly determines the nonlocal distortion function by requiring bouncing solutions in the beginning of the universe. Section~\ref{sec.4} investigates the evolution of anisotropies near the bounce. Section~\ref{sec.5} draws our conclusions and discussions.

\section{The Deser-Woodard nonlocal gravity}\label{sec.2}

We start with the action of the DW nonlocal gravity model as follows \cite{Deser:2007jk}:
\begin{equation}
\mathcal{S}=\frac{1}{16\pi}\int d^4x\sqrt{-g}R\left[1+f\left(\Box^{-1}R\right)\right]+\mathcal{S}_m\,,\label{action1}
\end{equation}
where $\mathcal{S}_m$ stands for the matter action. Here $f$ is an arbitrary function of the inverse scalar d'Alembertian acting on the Ricci scalar $R$. This function $f$, called the nonlocal distortion function, adds nonlocal modifications to the Einstein-Hilbert action. Note that we have assumed $c=G=1$ in this paper.

The nonlocal action \eqref{action1} can be localized by introducing an auxiliary scalar field
\begin{equation}
\Box^{-1}R\equiv\phi\,,
\end{equation}
and a Lagrange multiplier $\xi$ \cite{Nojiri:2007uq}. The action \eqref{action1} then can be rewritten as
\begin{align}
\mathcal{S}&=\frac{1}{16\pi}\int d^4x\sqrt{-g}\left[R\left(1+f\left(\phi\right)\right)+\xi\left(\Box\phi-R\right)\right]+\mathcal{S}_m\nonumber\\
&=\frac{1}{16\pi}\int d^4x\sqrt{-g}\left[R\left(1+f\right)-\partial_\mu\xi\partial^\mu\phi-\xi R\right]+\mathcal{S}_m\,,\label{action2}
\end{align}
where the second equality is deduced with an integration by part.

The equations of motion can be derived by varying the action \eqref{action2} with respect to $\xi$, $\phi$, and the metric $g_{\mu\nu}$. Varying the action with respect to the Lagrange multiplier $\xi$ leads to 
\begin{equation}
\Box\phi=\frac{1}{\sqrt{-g}}\partial_\mu\left(\sqrt{-g}\partial^{\mu}\phi\right)=R\,.\label{boxphi}
\end{equation} 
The variation with respect to the auxiliary scalar field $\phi$ gives
\begin{equation}
\Box\xi=-R\frac{df}{d\phi}\,.\label{boxxi}
\end{equation}
Finally, varying the action with respect to the metric $g_{\mu\nu}$ leads to the modified Einstein field equation
\begin{equation}
G_{\mu\nu}+\Delta G_{\mu\nu}=8\pi T_{\mu\nu}\,,\label{modifiedeinsteineq}
\end{equation}
where
\begin{align}
\Delta G_{\mu\nu}=&\,\left(G_{\mu\nu}+g_{\mu\nu}\Box-\nabla_\mu\nabla_\nu\right)\left(f-\xi\right)\nonumber\\&+\frac{1}{2}g_{\mu\nu}\partial_\rho\xi\partial^\rho\phi-\partial_{(\mu}\xi\partial_{\nu)}\phi\,,
\end{align}
where $G_{\mu\nu}$ and $T_{\mu\nu}$ stand for the Einstein tensor and the energy momentum tensor, respectively. 

It can be noted from the field equation \eqref{modifiedeinsteineq} that all the nonlocal modifications are encoded in the correction term $\Delta G_{\mu\nu}$. In the original context, the function $f$ was determined by requiring to generate the late-time acceleration of the universe without introducing dark energy in the matter sector \cite{Deffayet:2009ca}. See also \cite{Vernov:2012nr,Elizalde:2012ja} for the reconstruction procedures of $f$ for alternative cosmological evolutions. In the next section, we will instead reconstruct the function $f$ by requiring a bouncing solution in the very early universe. We will demonstrate that a primordial bouncing cosmology can be realized in the DW nonlocal gravity.

\section{Bouncing cosmology}\label{sec.3}

In order to construct a primordial bouncing solution, we consider a homogeneous, isotropic, and spatially flat universe which can be described by the flat Friedmann-Lema\^itre-Robertson-Walker (FLRW) metric. The line element reads
\begin{equation}
ds^2=-dt^2+a(t)^2dx_idx^i\,,\label{flrw}
\end{equation}
where the scale factor $a(t)$ is a function of cosmic time $t$. If we assume that the energy momentum tensor is governed by a perfect fluid with energy density $\rho$ and pressure $p$, the $(0,0)$ and $(i,j)$ components of the field equation \eqref{modifiedeinsteineq} can be written as
\begin{align}
3H^2\left(1+f-\xi\right)&+3H\left(\dot{f}-\dot{\xi}\right)-\frac{1}{2}\dot{\xi}\dot{\phi}=8\pi\rho\,,\label{eq009}\\
\left(3H^2+2\dot{H}\right)&\left(1+f-\xi\right)+\frac{1}{2}\dot{\xi}\dot{\phi}\nonumber\\&+\left(\frac{d^2}{dt^2}+2H\frac{d}{dt}\right)\left(f-\xi\right)+8\pi p=0\,,\label{eq100}
\end{align}
where the dot denotes $d/dt$ and $H\equiv\dot{a}/a$ is the Hubble function. Eliminating $\dot{\xi}\dot{\phi}$ terms, one can combine Eqs.~\eqref{eq009} and \eqref{eq100} to get \cite{Deffayet:2009ca}
\begin{equation}
\ddot{F}+5H\dot{F}+\left(6H^2+2\dot{H}\right)\left(1+F\right)=8\pi\left(\rho-p\right)\,,\label{reconeq}
\end{equation}
where $F\equiv f-\xi$. For any given cosmological evolution $a(t)$, one can in principle reconstruct the distortion function $f$ according to Eq.~\eqref{reconeq}.

Within the FLRW spacetime \eqref{flrw}, the equations for scalar fields $\phi$ and $\xi$, Eqs.~\eqref{boxphi} and \eqref{boxxi}, can be written as
\begin{align}
\ddot{\phi}+3H\dot{\phi}+6\left(\dot{H}+2H^2\right)&=0\,,\label{phieq}\\
\ddot{\xi}+3H\dot{\xi}-6\left(\dot{H}+2H^2\right)\frac{df}{d\phi}&=0\,,\label{xiieq}
\end{align}
respectively.

\subsection{The primordial bounce}

We assume that the expansion of the universe started with a primordial bounce. For the simplest bouncing solution, we suppose that the scale factor $a(t)$ can be written as  
\begin{equation}
a(t)=a_be^{h_1t^2/2}\,,\qquad H(t)=h_1t\,,\label{generalbounce}
\end{equation}
where $a_b$ is the minimum size of the scale factor at the bounce, and $h_1$ is a positive constant. We will focus on the positive branch of the cosmic time since only the expanding phase of the universe with a positive Hubble function will be considered in this work. See Refs.~\cite{Koshelev:2012qn,Biswas:2012bp} for the investigation of bouncing solutions in different classes of nonlocal gravity models.

It should be emphasized that a general bouncing solution is not necessarily in the form of Eq.~\eqref{generalbounce}. However, without loss of generality, the scale factor of most physical bouncing solutions can be approximated as
\begin{equation}
a(t)\approx a_b\left(1+\frac{h_1}{2}t^2\right)\,,\qquad H(t)\approx h_1t\,.\label{bouncea}
\end{equation}
Therefore, we will assume a Hubble function linear in time to characterize the bouncing solution in this paper.

Furthermore, we shall consider a primordial bounce occurring at the very early universe prior to the reheating epoch. Since the bouncing solution is introduced to replace the big bang singularity, the epoch that we consider here for the cosmological solution should be even before the inflationary phase. In this respect, the spacetime can be effectively regarded as a vacuum universe $(T_{\mu\nu}=0)$. Inserting Eq.~\eqref{generalbounce} into Eqs.~\eqref{reconeq}, \eqref{phieq}, and \eqref{xiieq}, we get
\begin{align}
F''+5\tau F'+2\left(1+3\tau^2\right)\left(1+F\right)&=0\,,\label{32}\\
\phi''+3\tau\phi'+6\left(1+2\tau^2\right)&=0\,,\label{33}\\
\xi''+3\tau\xi'-6\left(1+2\tau^2\right)\frac{df}{d\phi}&=0\,.\label{34}
\end{align}
Note that we have introduced a dimensionless time variable $\tau\equiv\sqrt{h_1}t$ for the sake of convenience. In the above equations, the prime denotes $d/d\tau$. The exact solutions to Eqs.~\eqref{32} and \eqref{33} read
\begin{align}
F(\tau)=&\,e^{-\tau^2}\left[A_1+A_2\,\textrm{erf}\left(\frac{\tau}{\sqrt{2}}\right)\right]-1\,,\label{solFacc}\\
\phi(\tau)=B_1&+B_2\,\textrm{erf}\left(\sqrt{\frac{3}{2}}\tau\right)\nonumber\\&-\left[2+\,_2F_2\left(1,1;\frac{3}{2},2;-\frac{3\tau^2}{2}\right)\right]\tau^2\,,\label{solphiacc}
\end{align}
where $A_1$, $A_2$, $B_1$, and $B_2$ are integration constants, and $\textrm{erf}(x)$ and $_pF_q(...)$ are the error function and the generalized hypergeometric function, respectively. From Eq.~\eqref{solphiacc}, we can derive
\begin{equation}
\phi'=\sqrt{\frac{6}{\pi}}B_2e^{-3\tau^2/2}-4\tau-\sqrt{\frac{2\pi}{3}}e^{-3\tau^2/2}\textrm{erfi}\left(\sqrt{\frac{3}{2}}\tau\right)\,,
\end{equation}
where $\textrm{erfi}(x)$ is the imaginary error function. We assume $B_2\le0$ such that $\phi'\le0$ for $\tau\ge0$. The reason of making this assumption will be clearer later when initial conditions are imposed. 

To reconstruct the distortion function $f=F+\xi$, we need to find $\xi(\tau)$. Therefore, we numerically solve Eq.~\eqref{eq009}
\begin{equation}
\xi'\phi'=6\tau^2\left(1+F\right)+6\tau F'\,,\label{eqxi}
\end{equation}
which turns out to be a first order differential equation of $\xi(\tau)$. After obtaining $\xi(\tau)$, we can use $f(\tau)=F(\tau)+\xi(\tau)$ and $\phi(\tau)$ to get $f(\phi)$, completing the reconstruction of the distortion function.

\subsection{Initial conditions}

In Ref.~\cite{Deffayet:2009ca}, the distortion function generating the $\Lambda$CDM cosmology is fitted into
\begin{equation}
f(\phi)=0.245\left[\tanh\left(0.35Y+0.032Y^2+0.003Y^3\right)-1\right]\,,
\end{equation}
where $Y\equiv\phi+16.5$. 
The auxiliary scalar field $\phi$ is negative in the $\Lambda$CDM cosmology. Also, from the qualitative behavior of $f(\phi)$ presented in Ref.~\cite{Deffayet:2009ca}, one can see that $\phi'$ is always negative, hence $\phi$ decreases in time. 

We will assume $\phi'\le0$ in our bouncing solutions, and that is why we required $B_2\le0$ in the previous subsection. In addition, we demand that at a certain time $\tau_i$ in the early universe (while $\tau_i$ is still far away from the bounce), the distortion function satisfies the following two conditions: $f\lesssim0$ and $df/d\phi\gtrsim0$. We also require that $\phi(\tau_i)=0$ by fixing the integration constant $B_1$ properly. In this regard, the bounce ($\tau=0$) would take place with a positive $\phi$. See Figures~\ref{f1}, \ref{f2}, and \ref{f3} in the next subsection.

The first condition $f\lesssim0$ can be achieved by fixing the integration constant $\xi(\tau_i)$ when solving Eq.~\eqref{eqxi}. More precisely, we can choose the value of $\xi(\tau_i)$ such that $f(\tau_i)=F(\tau_i)+\xi(\tau_i)\lesssim0$. Notice that $F(\tau)\rightarrow-1$ when $\tau\gg1$.

In order to reconcile the solution with the second condition $df/d\phi\gtrsim0$, we calculate $df/d\phi=f'/\phi'$ according to Eqs.~\eqref{solFacc}, \eqref{solphiacc}, and \eqref{eqxi}. When $\tau\gg1$, the result can be approximated as
\begin{equation}
\frac{df}{d\phi}\approx\frac{1}{8}\left(A_1+A_2\right)e^{-\tau^2}\,.
\end{equation}
Therefore, as long as $A_1+A_2>0$, the second condition $df/d\phi\gtrsim0$ can be satisfied at $\tau_i$.

\subsection{Reconstruction of $f(\phi)$}

\begin{figure}[t]
\includegraphics[scale=0.62]{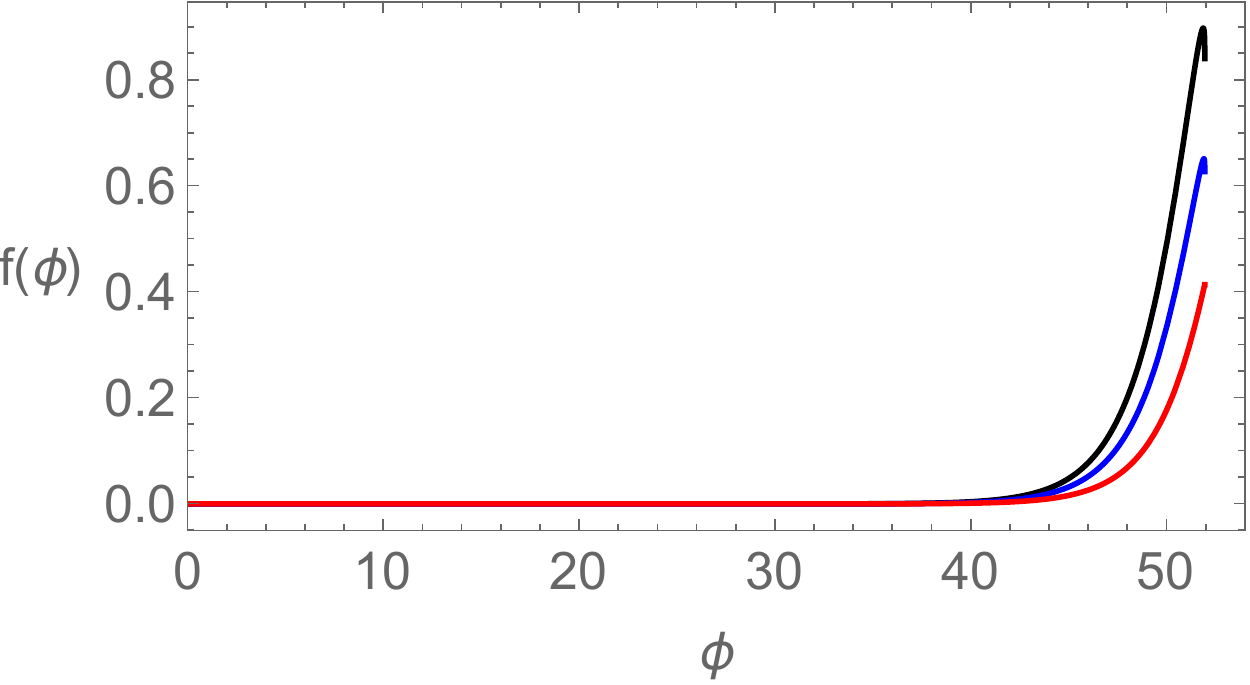}
\caption{\label{f1}The distortion function $f=F+\xi$ as a function of $\phi$ is shown. Here we fix $B_2=-0.1$ and $A_1=1$. The black, blue, and red curves correspond to $A_2=2$, $A_2=1$, and $A_2=0$, respectively. The initial condition $\xi=0.999$ is chosen at $\tau_i=5$. The bounce takes place at $\phi\approx50$ where the distortion function deviates significantly from zero. As time goes on, the field $\phi$ decreases and the distortion function approaches zero.}
\end{figure}

\begin{figure}[t]
\includegraphics[scale=0.62]{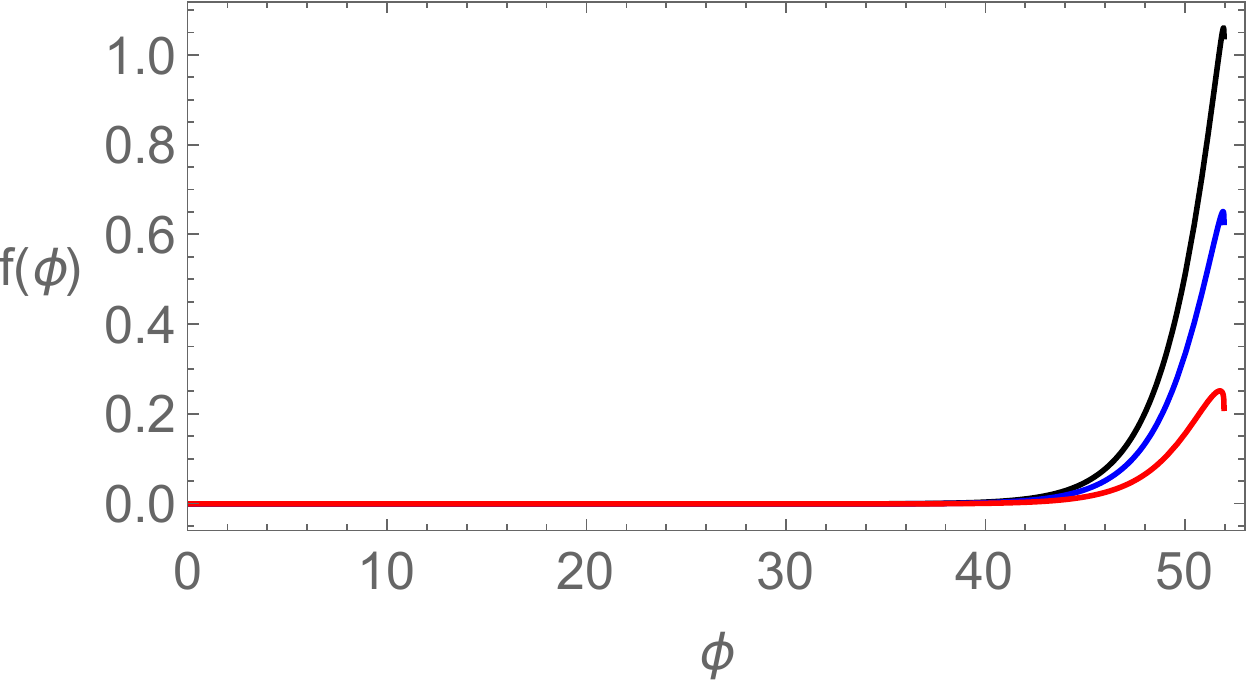}
\caption{\label{f2}The distortion function $f=F+\xi$ as a function of $\phi$ is shown. Here we fix $B_2=-0.1$ and $A_2=1$. The black, blue, and red curves correspond to $A_1=2$, $A_1=1$, and $A_1=0$, respectively. The initial condition $\xi=0.999$ is chosen at $\tau_i=5$. The bounce takes place at $\phi\approx50$ where the distortion function deviates significantly from zero. As time goes on, the field $\phi$ decreases and the distortion function approaches zero.}
\end{figure}

\begin{figure}[t]
\includegraphics[scale=0.62]{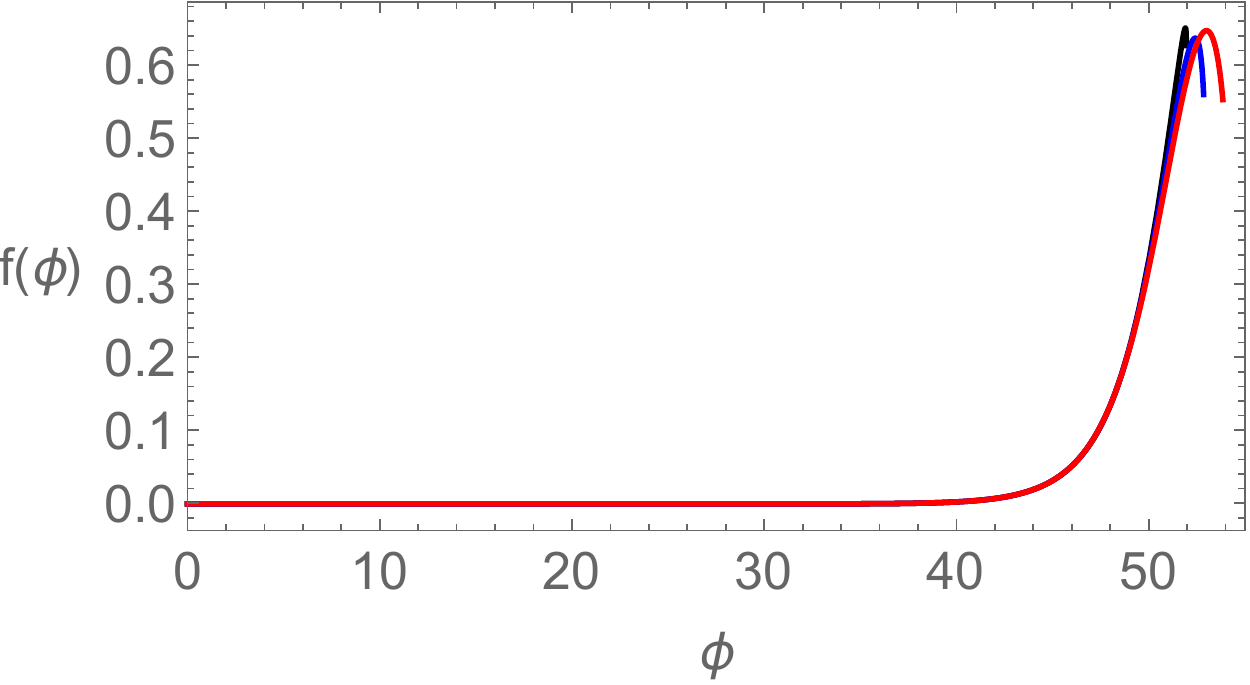}
\caption{\label{f3}The distortion function $f=F+\xi$ as a function of $\phi$ is shown. Here we fix $A_1=A_2=1$. The black, blue, and red curves correspond to $B_2=-0.1$, $B_2=-1$, and $B_2=-2$, respectively. The initial condition $\xi=0.999$ is chosen at $\tau_i=5$. The bounce takes place at $\phi\approx50$ where the distortion function deviates significantly from zero. As time goes on, the field $\phi$ decreases and the distortion function approaches zero.}
\end{figure}

After imposing appropriate initial conditions mentioned in the previous subsection, we obtain the distortion function $f(\phi)$ numerically and present the results in Figures~\ref{f1}, \ref{f2}, and \ref{f3}. 

In Figure~\ref{f1}, we fix $A_1$ and $B_2$, and show the distortion function with different values of $A_2$. In Figure~\ref{f2}, on the other hand, we fix $A_2$ and $B_2$. The distortion functions with different values of $A_1$ are exhibited. Finally, we fix $A_1$ and $A_2$, and see how the distortion function changes when $B_2$ varies. According to these figures, it can be seen that the distortion function in each case approaches zero as $\tau$ increases ($\phi$ decreases). It deviates from zero near the bounce when $\tau\rightarrow0$. It is therefore shown that by choosing the integration constants properly, the value of $f(\phi)$ and its derivative $df/d\phi$ approach zero after the bounce. This means that the distortion function of our model could be smoothly connected to that describing the standard $\Lambda$CDM cosmology.

\section{Evolution of anisotropies}\label{sec.4}
In several bouncing scenarios, the anisotropy would grow as $a^{-6}$ towards the bounce hence cannot be neglected \cite{Lehners:2008vx,Battefeld:2014uga,Belinsky:1970ew}. In this section, we will investigate how the anisotropy evolves near the bounce in the DW model. Considering a metric of the Bianchi I type
\begin{equation}
ds^2=-dt^2+a(t)^2\left[e^{2\beta_x(t)}dx^2+e^{2\beta_y(t)}dy^2+e^{2\beta_z(t)}dz^2\right]\,,\label{kasner}
\end{equation}
where the shear $\beta_i$ satisfies the condition $\sum_i\beta_i=0$, the field equation \eqref{modifiedeinsteineq} gives
\begin{align}
\left(3H^2-\sigma\right)(1+F)+3H\dot{F}&-\frac{1}{2}\dot{\xi}\dot{\phi}=8\pi\rho\,,\label{kasner00}\\
-\left(12H^2+6\dot{H}+2\sigma\right)(1+F)&-3\ddot{F}-9H\dot{F}-\dot{\xi}\dot{\phi}\nonumber\\=&\,8\pi(-\rho+3p)\,,\label{kasnertrace}
\end{align}
where $\sigma\equiv\sum_i\dot{\beta}_i^2/2$ stands for the shear density of anisotropies. After eliminating $\dot{\xi}\dot{\phi}$ terms, Eq.~\eqref{reconeq} is recovered. Therefore, the solution for $F(\tau)$ given in Eq.~\eqref{solFacc} is still valid after including anisotropies when the bouncing ansatz \eqref{generalbounce} is considered. 

Most importantly, if we use Eqs.~\eqref{kasner00}, \eqref{kasnertrace}, and the two scalar field equations \eqref{boxphi} and \eqref{boxxi}, we obtain
\begin{equation}
\dot\sigma+\left(6H+\frac{2\dot{F}}{1+F}\right)\sigma=0\,,\label{rhobeta1}
\end{equation}
which can be solved to get $\sigma\propto (1+F)^{-2}a^{-6}$. In GR ($\dot{F}=0$), the shear density grows as $\sigma\propto a^{-6}$ towards the bounce. In the DW gravity, however, the presence of nonlocal terms tends to suppress the growth of the shear density. According to Eqs.~\eqref{generalbounce} and \eqref{solFacc}, the function $F$ can be estimated as $F\approx a^{-2}$ near the bounce as long as $A_1$ is not zero. In this regard, the shear density in the DW bouncing model still grows, but in a milder way, i.e., $\sigma\propto a^{-2}$.

\section{Conclusions}\label{sec.5}

We have considered primordial bouncing cosmology in the framework of the Deser-Woodard nonlocal gravity model. In the presence of the nonlocal modifications encoded in the distortion function $f$, the big bang singularity can be replaced with a bouncing scenario by including a distortion function appropriately. In this regard, the curvature of the spacetime remains finite and the curvature singularity is removed. The bouncing solution is supported by the nonlocal correction terms in the gravitational action in the sense that the null convergence condition is violated while the null energy condition of the standard energy-momentum tensor is not necessarily violated. In addition, we have shown that the growth of anisotropies towards the bounce is suppressed by the nonlocal effects ($\sigma\propto a^{-2}$). Actually, according to the perturbation equations for vector modes and tensor modes \cite{Amendola:2019fhc,Koivisto:2008dh}, it can be proven that the vector modes and tensor modes are linearly stable at the bounce as long as $A_1\ne0$. The three sets of bouncing solutions corresponding to the different reconstructions of the distortion function $f$ are presented in  Figures~\ref{f1}, \ref{f2}, and \ref{f3}. In each case, the initial conditions are chosen such that the distortion function and its first order derivative approach zero after the bounce and hence there is no modification to the Einstein-Hilbert action afterwards. This construction is in contrast to the original determination of $f$ in order to generate the late-time acceleration \cite{Deffayet:2009ca}. Given that the original DW model \cite{Deser:2007jk} as a solution for the late-time acceleration without dark energy is ruled out by the solar system constraints \cite{Belgacem:2018wtb}, it may be instead considered as a bouncing cosmology model to solve the initial singularity problem. It would be also interesting to see if bouncing solutions are possible in the new DW model  \cite{Deser:2019lmm}.
 
Although the reconstruction in this work is performed in a vacuum spacetime, the inclusion of normal matter, such as radiation, should not change our conclusion. Firstly, we consider a regime which is still far away from the bounce such that the standard Friedmann equation $H^2\propto\rho$ is still valid. Since $F$ grows as $a^{-2}$ towards the bounce according to the estimate in Section~\ref{sec.4}, the appearance of nonlocal effects will modify the Friedmann equation through the $H^2F$ term, which grows faster than normal matter (see Eq.~\eqref{eq100}). Hence, the nonlocal modifications will eventually dominate the normal matter field. Secondly, despite the normal matter field grows towards the bounce, its quantity acquires a maximum value when $a=a_b$. The shear density $\sigma$ grows even slower than radiation and pressureless particles. Therefore, they can be regarded as inhomogeneous terms in the differential equations, e.g., Eq.~\eqref{reconeq}. A particular solution for the differential equations turns out to be sub-dominant compared with the homogeneous solutions. Moreover, with the field equations and the bouncing ansatz, the distortion function can always be reconstructed by solving the corresponding differential equations. Therefore, this conclusion is not expected to change 
in the presence of normal matter fields.

\section*{Acknowledgments}
CYC and PC are supported by Ministry of Science and Technology (MOST), Taiwan, through grant No.107-2119-M-002-005, Leung Center for Cosmology and Particle Astrophysics (LeCosPA) of National Taiwan University, and Taiwan National Center for Theoretical Sciences (NCTS). PC is in addition supported by US Department of Energy under Contract No. DE-AC03-76SF00515. 
SP is supported by the European Research Council under the European Union's Seventh Framework Programme 
(FP7/2007-2013)/ERC Grant No. 617656, ``Theories and Models of the Dark Sector: Dark Matter, Dark Energy and Gravity".
SP also acknowledges the Taiwan Travel Fellowship received from the Czech Academy of Sciences (CAS), which supported her visit to LeCosPA and the initiation of the current project. SP is grateful for the hospitality provided by LeCosPA.

\end{document}